\def~{\penalty\@M\kern3pt}

\def\thebibliography#1{\section*{References}\small\list
  {\arabic{enumi}.}{\settowidth\labelwidth{#1.}\leftmargin\labelwidth
    \advance\leftmargin\labelsep
    \usecounter{enumi}}
    \def\newblock{\hskip .11em plus .33em minus -.07em}
    \sloppy
    \sfcode`\.=1000\relax}

\def\ds@citeauthoryear{\def\thebibliography##1{\section*{References}%
    \small\list{}{\settowidth\labelwidth{}\leftmargin\parindent
    \itemindent=-\parindent
    \labelsep=\z@
    \usecounter{enumi}}%
    \def\newblock{\hskip .11em plus .33em minus -.07em}%
    \sloppy
    \sfcode`\.=1000\relax}%
    \def\@cite##1{##1}%
    \def\@lbibitem[##1]##2{\item[]\if@filesw
      {\def\protect####1{\string ####1\space}\immediate
    \write\@auxout{\string\bibcite{##2}{##1}}}\fi\ignorespaces}}%

\newif\if@envcountreset\@envcountresetfalse
\def\ds@envcountreset{\@envcountresettrue}

\def\@mbi{cmmib10}
\def\@ptsize{0} \@namedef{ds@11pt}{\def\@ptsize{1}}
\@namedef{ds@12pt}{\def\@ptsize{2}}
\def\ds@twoside{\@twosidetrue \@mparswitchtrue}
\def\ds@draft{\overfullrule
5pt}
\@options

\ds@twoside

\def\@normalsize{\@setsize\normalsize{12pt}\xpt\@xpt
\abovedisplayskip=3 mm plus6pt minus 4pt
\belowdisplayskip=3 mm plus6pt minus 4pt
\abovedisplayshortskip=0mm plus6pt minus 2pt
\belowdisplayshortskip=2 mm plus4pt minus 4pt}

\mathchardef\Gamma="0100
\mathchardef\Delta="0101
\mathchardef\Theta="0102
\mathchardef\Lambda="0103
\mathchardef\Xi="0104
\mathchardef\Pi="0105
\mathchardef\Sigma="0106
\mathchardef\Upsilon="0107
\mathchardef\Phi="0108
\mathchardef\Psi="0109
\mathchardef\Omega="010A

\def\small{\@setsize\small{11pt}\ixpt\@ixpt
\abovedisplayskip=2.5 mm plus5pt minus 3pt
\belowdisplayskip=2.5 mm plus5pt minus 3pt
\abovedisplayshortskip=0mm plus6pt minus 2pt
\belowdisplayshortskip=2 mm plus4pt minus 4pt
\def\@listi{\leftmargin\leftmargini\topsep 4pt plus 2pt minus 2pt}}

\def\footnotesize{\@setsize\footnotesize{11pt}\ixpt\@ixpt
\abovedisplayskip=2.5 mm plus5pt minus 3pt
\belowdisplayskip=2.5 mm plus5pt minus 3pt
\abovedisplayshortskip=0mm plus6pt minus 2pt
\belowdisplayshortskip=2 mm plus4pt minus 4pt
\def\@listi{\leftmargin\leftmargini\topsep 4pt plus 2pt minus 2pt}}

\def\scriptsize{\@setsize\scriptsize{8.4pt}\viipt\@viipt}

\def\tiny{\@setsize\tiny{6pt}\vpt\@vpt}
\def\large{\@setsize\large{13.2pt}\xipt\@xipt}
\def\Large{\@setsize\Large{14.4pt}\xiipt\@xiipt}
\def\LARGE{\@setsize\LARGE{16.8pt}\xivpt\@xivpt}
\def\huge{\@setsize\huge{22pt}\xxpt\@xxpt}
\def\Huge{\@setsize\Huge{30pt}\xxvpt\@xxvpt}
\@normalsize

\if@twoside
\else \oddsidemargin 63pt \evensidemargin 63pt
\fi
\skip\footins 9pt plus 4pt minus 2pt
\@beginparpenalty-\@lowpenalty\@endparpenalty -\@lowpenalty\@itempenalty
-\@lowpenalty

\def\@sect#1#2#3#4#5#6[#7]#8{\ifnum #2>\c@secnumdepth
  \def\@svsec{}\else
  \refstepcounter{#1}\edef\@svsec{\csname the#1\endcsname\quad }\fi
  \@tempskipa #5\relax
   \ifdim \@tempskipa>\z@
  \begingroup #6\relax
  \noindent{\hskip #3\relax\@svsec}{\interlinepenalty \@M #8\par}
  \endgroup
    \csname #1mark\endcsname{#7}\addcontentsline
   {toc}{#1}{\ifnum #2>\c@secnumdepth \else
            \protect\numberline{\csname the#1\endcsname}\fi
          #7}\else
  \def\@svsechd{#6\hskip #3\@svsec #8\csname #1mark\endcsname
            {#7}\addcontentsline
             {toc}{#1}{\ifnum #2>\c@secnumdepth \else
               \protect\numberline{\csname the#1\endcsname}\fi
             #7}}\fi
  \@xsect{#5}}

\def\part{\par \addvspace{4ex} \@afterindentfalse \secdef\@part\@spart}

\def\@part[#1]#2{\ifnum \c@secnumdepth >\m@ne \refstepcounter{part}
\addcontentsline{toc}{part}{\thepart \hspace{1em}#1}\else
\addcontentsline{toc}{part}{#1}\fi { \parindent 0pt \raggedright
 \ifnum \c@secnumdepth >\m@ne \Large \bf
 Part\thepart\par\nobreak\fi\huge
\bf #2\markboth{}{}\par } \nobreak \vskip 3ex \@afterheading }

\def\@spart#1{{\parindent 0pt \raggedright
 \huge \bf
 #1\par} \nobreak \vskip 3ex \@afterheading }

\def\section{\@startsection {section}{1}{\z@}{-18pt plus -4pt minus
-4pt}{12pt plus 4pt minus 4pt}{\Large\bf\boldmath
\pretolerance=10000\relax\rightskip=0pt plus8em}}
\def\subsection{\@startsection{subsection}{2}{\z@}{-18pt plus-4pt minus
 -4pt}{8pt plus 4pt minus 4pt}{\normalsize\bf\boldmath
\pretolerance=10000\relax\rightskip=0pt plus8em}}
\def\subsubsection{\@startsection{subsubsection}{3}{\z@}{-18pt plus-4pt
 minus -4pt}{-0.5em plus -.22em minus -0.1em}{\normalsize\bf\boldmath}}
\def\paragraph{\@startsection{paragraph}{4}{\z@}{-12pt plus -4pt minus
 -4pt}{-0.5em plus -.22em minus -0.1em}{\normalsize\it}}
\def\subparagraph#1{\typeout{LLNCS Warning: You should not use
\protect\subparagraph \space in this style.}\vskip0.5cm
You should not use $\backslash${\tt subparagraph} in this
style.\vskip0.5cm}

\def\appendix{\par
 \setcounter{section}{0}
 \setcounter{subsection}{0}
 \def\thesection{\Alph{section}}}

\leftmargin\leftmargini \labelwidth\leftmargini
\advance\labelwidth-\labelsep
\def\@listi{\leftmargin\leftmargini}
\def\@listii{\leftmargin\leftmarginii
 \labelwidth\leftmarginii\advance\labelwidth-\labelsep
 \topsep 0pt plus 1pt}
\def\@listiii{\leftmargin\leftmarginiii
 \labelwidth\leftmarginiii\advance\labelwidth-\labelsep
 \topsep 0pt plus 1pt}
\def\@listiv{\leftmargin\leftmarginiv
 \labelwidth\leftmarginiv\advance\labelwidth-\labelsep
 \topsep 0pt plus 1pt}
\def\@listv{\leftmargin\leftmarginv
 \labelwidth\leftmarginv\advance\labelwidth-\labelsep
 \topsep 0pt plus 1pt}
\def\@listvi{\leftmargin\leftmarginvi
 \labelwidth\leftmarginvi\advance\labelwidth-\labelsep
 \topsep 0pt plus 1pt}

\def\theenumi{\arabic{enumi}}

\def\theenumii{\alph{enumii}}
\def\p@enumii{\theenumi}

\def\theenumiii{\roman{enumiii}}
\def\p@enumiii{\theenumi(\theenumii)}

\def\p@enumiv{\p@enumiii\theenumiii}

\def\verse{\let\\=\@centercr
 \list{}{\itemsep\z@ \itemindent -1.5em\listparindent \itemindent
 \rightmargin\leftmargin\advance\leftmargin 1.5em}\item[]}

\def\descriptionlabel#1{\hspace\labelsep \bf #1}
\def\description{\list{}{\labelwidth\z@ \itemindent-\leftmargin
 \let\makelabel\descriptionlabel}}

\def\titlepage{\@restonecolfalse\if@twocolumn\@restonecoltrue\onecolumn
 \else \newpage \fi \thispagestyle{empty}\c@page\z@}
\def\endtitlepage{\if@restonecol\twocolumn \else \newpage \fi}

\tabbingsep \labelsep

\skip\@mpfootins = \skip\footins
\newcounter{part}
\newcounter {section}
\newcounter {subsection}[section]
\newcounter {subsubsection}[subsection]
\newcounter {paragraph}[subsubsection]

\def\thepart{\Roman{part}}
\def\thesection {\arabic{section}}

\def\@pnumwidth{1.55em}
\def\@tocrmarg {2.55em}
\def\@dotsep{4.5}
\def\tableofcontents{\section*{Table of Contents}
 \@starttoc{toc}}
\def\l@part#1#2{\addpenalty{\@secpenalty}
 \addvspace{2.25em plus 1pt} \begingroup
 \@tempdima 3em \parindent \z@ \rightskip \@pnumwidth \parfillskip
-\@pnumwidth
 {\Large \bf \leavevmode #1\hfil \hbox to\@pnumwidth{\hss #2}}\par
 \nobreak \endgroup}
\def\numberline#1{\advance\hangindent by\@tempdima%
\hbox to\@tempdima{\hss#1\enspace}}
\def\bf@dottedtocline#1#2#3#4#5{\ifnum #1>\c@tocdepth \else
  \vskip \z@ plus .2pt
  {\leftskip #2\relax \rightskip \@tocrmarg \parfillskip -\rightskip
    \parindent #2\relax\@afterindenttrue
   \interlinepenalty\@M
   \leavevmode
   \@tempdima #3\relax \advance\leftskip \@tempdima \hbox{}\hskip
   -\leftskip
{\bf#4}\nobreak\leaders\hbox{$\m@th \mkern \@dotsep mu.\mkern \@dotsep
       mu$}\hfill \nobreak \hbox to\@pnumwidth{\hfil\rm #5}\par}\fi}
\def\l@section{\vskip2mm\bf@dottedtocline{1}{0em}{1.7em}}
\def\l@subsection{\@dottedtocline{2}{1.7em}{2.3em}}
\def\l@subsubsection{\@dottedtocline{3}{4em}{2em}}
\def\l@paragraph{\@dottedtocline{4}{6em}{2em}}
\def\l@subparagraph{\@dottedtocline{5}{8em}{2em}}
\def\listoffigures{\section*{List of Figures\markboth
 {List of Figures}{List of Figures}}\@starttoc{lof}}
\def\l@figure{\@dottedtocline{1}{1.5em}{2.3em}}
\def\listoftables{\section*{List of Tables\markboth
 {List of Tables}{List of Tables}}\@starttoc{lot}}
\let\l@table\l@figure

\newif\if@restonecol
\def\theindex{\@restonecoltrue\if@twocolumn\@restonecolfalse\fi
\columnseprule \z@
\columnsep 35pt\twocolumn[\section*{Index}]
 \markboth{Index}{Index}\thispagestyle{plain}\parindent\z@
 \parskip\z@ plus .3pt\relax\let\item\@idxitem}
\def\@idxitem{\par\hangindent 40pt}

\def\endtheindex{\if@restonecol\onecolumn\else\clearpage\fi}

\def\footnoterule{\kern-3\p@\hrule width 2 true cm\kern 2.6\p@}

\long\def\@makefntext#1{\@setpar{\@@par\@tempdima \hsize
  \advance\@tempdima-1em\parshape \@ne 1em\@tempdima}\par
  \parindent 1em\noindent \hbox to \z@{\hss$^{\@thefnmark}$\ }#1}

\long\def\@makecaption#1#2{
 \vskip 10pt
 \setbox\@tempboxa\hbox{{\bf #1} #2}
 \ifdim \wd\@tempboxa >\hsize \unhbox\@tempboxa\par \else \hbox
to\hsize{\box\@tempboxa\hfil}
 \fi\vskip5pt}

\long\def\@caption#1[#2]#3{\addcontentsline{\csname
  ext@#1\endcsname}{#1}{\protect\numberline{\csname
  the#1\endcsname}{\ignorespaces #2}}\par
  \begingroup
    \@parboxrestore
    \@makecaption{\csname fnum@#1\endcsname}{\ignorespaces #3}\par
  \endgroup}

\newcounter{figure}
\def\thefigure{\@arabic\c@figure}

\def\fps@figure{htbp}
\def\ftype@figure{1}
\def\ext@figure{lof}
\def\fnum@figure{Fig.\thinspace\thefigure.}
\def\figure{\@ifnextchar[{\@yfigure}{\@xfigure}}
\def\@xfigure{\@float{figure}\small\rm}
\def\@yfigure[#1]{\@float{figure}[#1]\small\rm}
\def\endfigure{\vskip-5pt\end@float}
\@namedef{figure*}{\@dblfloat{figure}}
\@namedef{endfigure*}{\end@dblfloat}

\newcounter{table}
\def\thetable{\@arabic\c@table}
\def\fps@table{htbp}
\def\ftype@table{2}
\def\ext@table{lot}
\def\fnum@table{Table \thetable.}
\def\table{\@float{table}\small\rm}
\let\endtable\end@float
\@namedef{table*}{\@dblfloat{table}}
\@namedef{endtable*}{\end@dblfloat}

\newcounter{@inst}
\newcounter{@auth}

\def\institute#1{\gdef\@institute{#1}}

\def\institutename{\par
 \begingroup
 \parskip=\z@
 \parindent=\z@
 \setcounter{@inst}{1}%
 \def\and{\par\stepcounter{@inst}%
 \noindent$^{\the@inst}$\enspace\ignorespaces}%
 \setbox0=\vbox{\def\thanks##1{}\@institute}%
 \ifnum\c@@inst=1\relax
 \else
   \setcounter{footnote}{\c@@inst}%
   \setcounter{@inst}{1}%
   \noindent$^{\the@inst}$\enspace
 \fi
 \ignorespaces
 \@institute\par
 \endgroup}

\def\@thanks{}

\def\@fnsymbol#1{\ifcase#1\or\star\or{\star\star}\or{\star\star\star}%
   \or \dagger\or \ddagger\or
   \mathchar "278\or \mathchar "27B\or \|\or **\or \dagger\dagger
   \or \ddagger\ddagger\else\@ctrerr\fi\relax}

\def\subtitle#1{\gdef\@subtitle{#1}}
\def\@subtitle{}

\def\maketitle{\par
 \begingroup
 \parindent=\z@
 \def\thefootnote{\fnsymbol{footnote}}
 \if@twocolumn
 \twocolumn[\@maketitle]
 \else \newpage
 \global\@topnum\z@ \@maketitle \fi\thispagestyle{empty}\@thanks
 \endgroup
 \let\maketitle\relax
 \let\@maketitle\relax
 \gdef\@thanks{}\gdef\@author{}\gdef\@title{}\gdef\@subtitle{}%
 \let\thanks\relax}

\def\@maketitle{\newpage
 \begin{center}%
 {\LARGE \bf\boldmath
  \pretolerance=10000
  \@title \par}\vskip .8cm
\if!\@subtitle!\else {\Large \bf\boldmath
  \vskip -.65cm
  \pretolerance=10000
  \@subtitle \par}\vskip .8cm\fi
{\normalsize\rm\lineskip .5em
\@author\vskip.35cm}
 {\small\rm\institutename}
 \end{center}%
 }

\mark{{}{}}

\def\abstract{%
\list{}{\advance\topsep by0.35cm\relax\small\rm
 \leftmargin=1cm
 \labelwidth=\z@
 \listparindent=\z@
 \itemindent\listparindent
 \rightmargin\leftmargin}\item[\hskip\labelsep\bf Abstract.]}

\def\ps@headings{\def\@evenhead{\makebox[\textwidth]{\hfill
   \raisebox{2cm}[0pt][0pt]{\fbox{\LARGE\thepage}\hspace{-2cm}}}}%
\let\@oddhead\@evenhead
\def\@evenfoot{}
\let\@oddfoot\@evenfoot
\def\sectionmark##1{}%
\def\subsectionmark##1{}}
\def\ps@myheadings{\let\@mkboth\@gobbletwo
\def\@oddhead{\hbox{}\small\rm\rightmark \hfil\thepage}%
\def\@oddfoot{}\def\@evenhead{\small\rm\thepage\hfil
\leftmark\hbox {}}%
\def\@evenfoot{}\def\sectionmark##1{}\def\subsectionmark##1{}}

\def\today{\ifcase\month\or
 January\or February\or March\or April\or May\or June\or
 July\or August\or September\or October\or November\or December\fi
 \space\number\day, \number\year}

\ps@headings  \onecolumn
\if@twoside\else\fi

\def\newstytheorem#1#2#3{%
\@ifnextchar[{\che@othm{#1}{#2}{#3}}{\che@nthm{#1}{#2}{#3}}}
\def\che@nthm#1#2#3#4{%
\@ifnextchar[{\che@xnthm{#1}{#2}{#3}{#4}}{\che@ynthm{#1}{#2}{#3}{#4}}}
\def\che@xnthm#1#2#3#4[#5]{\expandafter
\@ifdefinable\csname #1\endcsname
{\@definecounter{#1}\if@envcountreset\@addtoreset{#1}{#5}\fi
\expandafter\xdef\csname the#1\endcsname{\@thmcounter{#1}}%
\global\@namedef{#1}{\che@thm{#1}{#4}{#2}{#3}}\global
\@namedef{end#1}{\@endtheorem}}}
\def\che@ynthm#1#2#3#4{\expandafter\@ifdefinable\csname #1\endcsname
{\@definecounter{#1}%
\expandafter\xdef\csname the#1\endcsname{\@thmcounter{#1}}%
\global\@namedef{#1}{\che@thm{#1}{#2}{#3}{#4}}\global
\@namedef{end#1}{\@endtheorem}}}
\def\che@othm#1#2#3[#4]#5{\expandafter\@ifdefinable
\csname#1\endcsname%
{\@@othm@{#1}{#2}{#3}{#4}{#5}\global\@namedef{end#1}{\@endtheorem}}}
\def\@@othm@#1#2#3#4#5{\if *#4{\global\@namedef{the#1}{\relax}
\global\@namedef{#1}{\@bthm@{}{#5}{#2}{#3}}}\else
{\global\@namedef{the#1}{\@nameuse{the#4}}
\global\@namedef{#1}{\che@thm{#4}{#5}{#2}{#3}}}\fi}
\def\che@thm#1#2#3#4{\@ifnextchar({\@athm{#1}{#2}{#3}{#4}}{%
\@ifnextchar *{\@bthm{#1}{#2}{#3}{#4}}%
{\@@thm{#1}{#2}{#3}{#4}}}}
\def\@athm#1#2#3#4(#5){\@ifnextchar[{%
\@aythm{#1}{#2}{#3}{#4}{#5}}{\@axthm{#1}{#2}{#3}{#4}{#5}}}
\def\@axthm#1#2#3#4#5{\@@locthmlab{#1}{#5}%
\@begintheorem{#2}{#5}{#3}{#4}\ignorespaces}
\def\@aythm#1#2#3#4#5[#6]{\@@locthmlab{#1}{#5}%
\fuh@opargbegintheorem{#2}{#5}{#6}{#3}{#4}\ignorespaces}
\def\@bthm@#1#2#3#4{\@ifnextchar[{\@bythm{#1}{#2}{#3}{#4}}{%
\@bxthm{#1}{#2}{#3}{#4}}}
\def\@bthm#1#2#3#4*{\@ifnextchar[{\@bythm{#1}{#2}{#3}{#4}}{%
\@bxthm{#1}{#2}{#3}{#4}}}
\def\@bxthm#1#2#3#4{\if !#1!\relax\else\@@locthmlab{#1}{}\fi
\@@begintheorem{#2}{#3}{#4}\ignorespaces}
\def\@bythm#1#2#3#4[#5]{%
\@@opargbegintheorem{#2}{#3}{#4}{#5}\ignorespaces}
\def\@@locthmlab#1#2{\expandafter\def\csname the#1\endcsname{#2}
\let\@chetempa\protect\def\protect{\noexpand\protect\noexpand}%
\edef\@currentlabel{\csname p@#1\endcsname\csname the#1\endcsname}%
\let\protect\@chetempa}
\def\@@thm#1#2#3#4{\refstepcounter
    {#1}\@ifnextchar[{\che@ythm{#1}{#2}{#3}{#4}}{%
    \che@xthm{#1}{#2}{#3}{#4}}}
\def\che@xthm#1#2#3#4{\che@begintheorem{#2}{\csname
the#1\endcsname}{#3}{#4}\ignorespaces}
\def\che@ythm#1#2#3#4[#5]{\che@opargbegintheorem{#2}{\csname
       the#1\endcsname}{#5}{#3}{#4}\ignorespaces}
\def\che@begintheorem#1#2#3#4{#4\trivlist\item[\hskip\labelsep
#3#1\ts #2.]}
\def\che@opargbegintheorem#1#2#3#4#5{#5\trivlist
\item[\hskip\labelsep#4#1\ts #2\ #3.]}
\def\fuh@opargbegintheorem#1#2#3{\it \trivlist
      \item[\hskip \labelsep{\bf #1\ #2\ (#3).}]}
\def\@@begintheorem#1#2#3{#3\trivlist\item[\hskip\labelsep
#2#1.]}
\def\@@opargbegintheorem#1#2#3#4{#3\trivlist\item[\hskip\labelsep
#2#1 #4.]}

\newstytheorem{theorem}{\bf}{\it}{Theorem}[section]
\newstytheorem{example}{\it}{\rm}{Example}[section]
\newstytheorem{proposition}{\bf}{\it}[theorem]{Proposition}
\newstytheorem{corollary}{\bf}{\it}[theorem]{Corollary}
\newstytheorem{lemma}{\bf}{\it}[theorem]{Lemma}
\newstytheorem{proof}{\it}{\rm}[*]{Proof}
\newstytheorem{definition}{\bf}{\rm}[theorem]{Definition}
\newstytheorem{remark}{\it}{\rm}[*]{Remark}
\newstytheorem{exercise}{\it}{\rm}[theorem]{Exercise}
\newstytheorem{problem}{\it}{\rm}[theorem]{Problem}
\newstytheorem{solution}{\it}{\rm}[theorem]{Solution}
\newstytheorem{note}{\it}{\rm}[theorem]{Note}
\newstytheorem{question}{\it}{\rm}[theorem]{Question}

\def\squareforqed{\hbox{\rlap{$\sqcap$}$\sqcup$}}
\def\qed{\ifmmode\squareforqed\else{\unskip\nobreak\hfil
\penalty50\hskip1em\null\nobreak\hfil\squareforqed
\parfillskip=0pt\finalhyphendemerits=0\endgraf}\fi}

\def\bbbc{{\mathchoice {\setbox0=\hbox{$\displaystyle\rm C$}\hbox{\hbox
to0pt{\kern0.4\wd0\vrule height0.9\ht0\hss}\box0}}
{\setbox0=\hbox{$\textstyle\rm C$}\hbox{\hbox
to0pt{\kern0.4\wd0\vrule height0.9\ht0\hss}\box0}}
{\setbox0=\hbox{$\scriptstyle\rm C$}\hbox{\hbox
to0pt{\kern0.4\wd0\vrule height0.9\ht0\hss}\box0}}
{\setbox0=\hbox{$\scriptscriptstyle\rm C$}\hbox{\hbox
to0pt{\kern0.4\wd0\vrule height0.9\ht0\hss}\box0}}}}
\def\bbbq{{\mathchoice {\setbox0=\hbox{$\displaystyle\rm
Q$}\hbox{\raise
0.15\ht0\hbox to0pt{\kern0.4\wd0\vrule height0.8\ht0\hss}\box0}}
{\setbox0=\hbox{$\textstyle\rm Q$}\hbox{\raise
0.15\ht0\hbox to0pt{\kern0.4\wd0\vrule height0.8\ht0\hss}\box0}}
{\setbox0=\hbox{$\scriptstyle\rm Q$}\hbox{\raise
0.15\ht0\hbox to0pt{\kern0.4\wd0\vrule height0.7\ht0\hss}\box0}}
{\setbox0=\hbox{$\scriptscriptstyle\rm Q$}\hbox{\raise
0.15\ht0\hbox to0pt{\kern0.4\wd0\vrule height0.7\ht0\hss}\box0}}}}
\def\bbbt{{\mathchoice {\setbox0=\hbox{$\displaystyle\rm
T$}\hbox{\hbox to0pt{\kern0.3\wd0\vrule height0.9\ht0\hss}\box0}}
{\setbox0=\hbox{$\textstyle\rm T$}\hbox{\hbox
to0pt{\kern0.3\wd0\vrule height0.9\ht0\hss}\box0}}
{\setbox0=\hbox{$\scriptstyle\rm T$}\hbox{\hbox
to0pt{\kern0.3\wd0\vrule height0.9\ht0\hss}\box0}}
{\setbox0=\hbox{$\scriptscriptstyle\rm T$}\hbox{\hbox
to0pt{\kern0.3\wd0\vrule height0.9\ht0\hss}\box0}}}}
\def\bbbs{{\mathchoice
{\setbox0=\hbox{$\displaystyle     \rm S$}\hbox{\raise0.5\ht0\hbox
to0pt{\kern0.35\wd0\vrule height0.45\ht0\hss}\hbox
to0pt{\kern0.55\wd0\vrule height0.5\ht0\hss}\box0}}
{\setbox0=\hbox{$\textstyle        \rm S$}\hbox{\raise0.5\ht0\hbox
to0pt{\kern0.35\wd0\vrule height0.45\ht0\hss}\hbox
to0pt{\kern0.55\wd0\vrule height0.5\ht0\hss}\box0}}
{\setbox0=\hbox{$\scriptstyle      \rm S$}\hbox{\raise0.5\ht0\hbox
to0pt{\kern0.35\wd0\vrule height0.45\ht0\hss}\raise0.05\ht0\hbox
to0pt{\kern0.5\wd0\vrule height0.45\ht0\hss}\box0}}
{\setbox0=\hbox{$\scriptscriptstyle\rm S$}\hbox{\raise0.5\ht0\hbox
to0pt{\kern0.4\wd0\vrule height0.45\ht0\hss}\raise0.05\ht0\hbox
to0pt{\kern0.55\wd0\vrule height0.45\ht0\hss}\box0}}}}
\def\bbbz{{\mathchoice {\hbox{$\sf\textstyle Z\kern-0.4em Z$}}
{\hbox{$\sf\textstyle Z\kern-0.4em Z$}}
{\hbox{$\sf\scriptstyle Z\kern-0.3em Z$}}
{\hbox{$\sf\scriptscriptstyle Z\kern-0.2em Z$}}}}
\def\ts{\thinspace}

\def\typeset{\vfill\small\noindent
\par}

\def\enddocument{\par\typeset
\@checkend{document}\clearpage\begingroup
\if@filesw \immediate\closeout\@mainaux
\def\global\@namedef##1##2{}\def\newlabel{\@testdef r}%
\def\bibcite{\@testdef b}\@tempswafalse\makeatletter\input \jobname.aux
\if@tempswa \@warning{Label(s) may have changed. Rerun to get
cross-references right}\fi\fi\endgroup\deadcycles\z@\@@end}

\documentstyle{llncs}

\begin{document}

\begin{center}
{\LARGE \bf A Dynamical Simulation Facility for Hybrid Systems}  \\
\vspace{.15in}
{\large \bf Allen Back, John Guckenheimer, Mark Myers} \\
{\large \bf Center for Applied Mathematics, Cornell University}
\end{center}

\vspace{.2in}

\begin{abstract}
\noindent This paper establishes a general framework for describing
hybrid dynamical
systems which is particularly suitable for numerical simulation.  In this
context, the data structures
used to describe the sets and functions which comprise the dynamical system
are crucial since
they provide the link between a natural mathematical formulation of
a problem
and the correct application of standard numerical algorithms.
We describe a  partial implementation of the
design methodology and use this simulation tool for
a specific control problem in robotics as an illustration of the utility of
the approach for practical applications.

\end{abstract}
\vspace{.2in}

\section*{A Definition of Hybrid Systems}

The last several decades have witnessed an explosive development
in the theory of
dynamical systems,  much of which is oriented toward equations
of the form
\begin{equation}
\label{eq1}
      \dot{x} = F(x,p)
\end{equation}
where $F:V \! \subset \! {\bf R}^n \! \times \!
 {\bf R}^k \! \rightarrow \! {\bf R}^n$ is a smooth
map defined on an open, connected subset of Euclidean space and
possibly dependent upon
a vector of parameters, $p \! \in {\bf R}^k$.
However, many areas of application frequently involve hybrid systems:
dynamical systems which require a mixture of discrete and
continuously evolving events.
Natural examples of such situations are mechanical systems which
involve impacts,
control systems that switch between a variety of feedback strategies,
and vector fields
defined on manifolds described by several charts. In this note we
present a uniform approach
for representing hybrid systems which applies to a wide variety of
problems and, in addition, is
particularly well suited for computer implementation.  Simulation
tools which exploit the
approach that we describe are being developed in a computer program
called {\bf dstool} \cite{back1}.

Our theoretical objective may be viewed as the natural extension of
Equation (\ref{eq1}) in two
fundamental ways: First, we wish to generalize the underlying domain
of $F$, denoted $V$ above, to include
open sets composed of many components where the vector field may vary
discontinuously from
component to component. Second, we want to accomodate the situation
that discrete events
depending upon the phase space coordinates or time may occur in
the flow of the dynamical system.  It is worth pointing out that
these goals represent fundamental changes in
the theoretical character of the dynamics.  By satisfying the
first requirement, we greatly increase
the class of systems under study, but must generalize
the underlying theory that supports the simulation and
analysis of such systems.
The second extension demands a formulation that treats the discrete
event components, or some representation of them, as equal
members of the underlying phase space with the
continuous parts.

Assume the problem domain may be decomposed into the form:
\[
   V = \bigcup_{\alpha \in I} V_\alpha
\]
where $I$ is a finite {\em index set} and $V_\alpha$ is an open,
connected subset of ${\bf R}^n$.
We shall refer to each element in this union as a {\em chart}.  Each
chart has associated with it
a (possibly time-dependent) vector field, $f_\alpha : V_\alpha \!
\times {\bf R} \! \rightarrow \! {\bf R}^n$.
Notice that the charts are not required to be disjoint.  Moreover, for
$\alpha, \beta \in I$ we do not require continuity, or even agreement,
of the vector fields on the intersection set
$V_\alpha \bigcap V_\beta$.  We introduce further structure by
requiring that for each $\alpha \in I$, the
chart $V_\alpha$ must enclose a {\em patch},
an open subset $U_{\alpha}$
satisfying  $\overline{U}_\alpha \! \subset \! V_\alpha$. The
boundary
of $U_{\alpha}$ is assumed piecewise smooth and is
referred to as the
{\em patch boundary}. Together, the collection of charts and
patches is called an  {\em atlas}.

To implement this model,  a concrete representation of
the patch boundaries must be selected.
For each $\alpha \in I$ we define
a finite set  of {\em boundary functions},
$h_{\alpha,i} : V_{\alpha} \rightarrow R$,
$ i \in J_{\alpha}^{bf}$,  and
real numbers called {\em target values}, $C_{\alpha,i}$,
for $i \in J_\alpha^{\! \! \mbox{ \scriptsize bf}}$
that satisfy the condition: For $x \in V_\alpha$ where
$\alpha \in I \; $, we require

\begin{tabbing}
000000000\=000\=0000000000\=00000\= \kill
  \>  $x \in {U}_\alpha \;$ if and only if
   $ \; h_{\alpha,i}(x)-C_{\alpha,i} > 0 \;$
  for all $i \in J_\alpha^{\! \! \mbox{ \scriptsize bf}} \; \;$ .
\end{tabbing}
Thus, a patch is to be considered the domain on which
a collection of smooth functions are positive.
The boundary of a patch is assumed to lie within the set:
\[
\bigcup_{i \in J_\alpha^{\! \! \mbox{ \scriptsize bf}}} h^{-1}_{\alpha,i}
\left( \left \{C_{\alpha,i} \right \} \right )
\hspace{.3in} \mbox{for} \; \alpha \in I \; .
\]
We remark that the target values $C_{\alpha,i}$ may depend
on the initial point of the trajectory segment as well as
the vector of system parameters. This dependence enables
a change of state to occur after a specified
time has elapsed.

Since ultimately we work with floating point
arithmetic, we shall specify an explicit tolerance
for the values of
$h_{\alpha,i}$ so that the operational definition of
the boundary is given by:

\begin{tabbing}
000000\=0000000\=0000000000\=00000\= \kill
  \>  \>  $x \in \partial U_\alpha$ if
$|h_{\alpha,i}(x)-C_{\alpha,i}|<\epsilon_{\alpha,i}$
$\; $ for some $i \in J_\alpha^{\! \! \mbox{ \scriptsize bf}} \;$.
\end{tabbing}

Note, however, that points not in $\partial U_\alpha$ may
satisfy this condition, and it may need to be supplemented
with inequalities describing which portions of a boundary
are described by a level set of $h_{\alpha,i}$.
Conceptually, the evolution of the system is viewed
as a sequence of trajectory segments where the endpoint of
one segment is connected to the initial point of the next
by a transformation.
It follows that time may be divided into contiguous
periods, called {\em epochs}, separated by instances
where {\em transition functions} are applied at times
referred to as {\em events}.  The transition functions
are maps which send a point on the boundary of one
patch to a point in
another (not necessarily different) patch in the
atlas.   The ``best'' definition of the transition functions
is still not apparent.  Events only occur at ``exit''
points from the boundary of a patch, so it is not
essential that transition functions be defined at
other boundary points of a patch.  We define
subsets $S_{\alpha}$ of the patch boundaries, called
{\em transition sets} and transformations
$T_{\alpha} : S_{\alpha} \rightarrow V \times I$ satisfying
$\pi_{1}(T_{\alpha}(x)) \in \overline{U}{_{\pi_2(T_{\alpha}(x))}}$
where $\pi_{1},\pi_{2}$ are the natural projection of $V \times I$
onto its
factors. Thus $\pi_{1}(T_{\alpha})$ is the ``continuous'' part
and $\pi_{2}(T_{\alpha})$ is the ``discrete'' part of a transition
function (mapping indices).
We require further that there is a finite partition
$\bigcup W_{\alpha,j}$ of $S_\alpha$ with the properties:

\begin{tabbing}
000000000\=0000\=0000000000\=00000\= \kill
\> i) \> $\pi_2 \circ T_\alpha$ is constant on each $W_{\alpha,j}$\\
\> ii) \> \rule{0in}{.25in} $\phi_{\alpha,j} =
\pi_1 \circ T_\alpha \rule[-.1in]{0.005in}{.21in}_{ \; W_{ \! \alpha,j}}$
is the restriction
of a smooth \\
\> \> \rule{.4in}{0in}  mapping defined in a neighborhood
of $\overline{W}_{\alpha,j} \; $ . \\
\end{tabbing}

Figure (\ref{exa}) illustrates the definitions presented
so far.  Within this framework, an orbit in the flow of
a hybrid dynamical system which begins at a time $t_0$
and terminates at $t_f$ may be
completely described.  A {\em trajectory} for
Equation (\ref{eq1}) is a curve
$\gamma$ : $[t_0,t_f] \!  \rightarrow V \! \times \! I $
together with an
increasing sequence of real numbers
$t_{0} < t_{1} < \cdots < t_{m}= t_f$
that satisfies three properties:

\begin{itemize}
\item Each time interval $(t_{i}, t_{i+1})$ corresponds to an epoch and
there exists a designated $\alpha$ so that $\gamma(t)$ lies
entirely in $\overline{U}{_{\alpha}}  \! \times \! \{ \alpha \}$
for all $t \in (t_{i}, t_{i+1})$.

\item  For $t \in [t_{i}, t_{i+1})$ and the unique $\alpha$
specified above,
$t \rightarrow \pi_{1} \left ( \gamma(t) \right )$ is an integral
curve of the vector field $f_{\alpha}$.

\item $\! \! \! \! \! {\displaystyle \lim_{\; \; \;
t \rightarrow t_{i+1}^-} \! \! \! \pi_1 \left (\gamma(t) \right ) = y } $
exists, $ \; y \in S_\alpha \; $
and $T_{\alpha}(y) = \! \! \! {\displaystyle \lim_{\;
\; \;  t \rightarrow t_{i+1}^{^{+}}}} \gamma(t) $.
\end{itemize}

A hybrid system will be called {\em complete} if
every trajectory can be continued
for $t \! \rightarrow \! \infty$.  This implies that every point
of $\partial U_\alpha$ reached by a trajectory lies
in the transition set $S_\alpha$ of $U_\alpha$.

%
%
%
%

One of our goals in the definition of hybrid systems has been to
retain the applicability of as many ``standard''
numerical algorithms as possible. Thus,
a design objective has been to ensure that algorithms for tasks such as
numerical integration and root finding can be used without modification.
This explains the seemingly pedantic distinction that we make between
charts and patches. Consider the task of locating the intersection
of a trajectory segment with a patch boundary. If an integration step
carries a trajectory outside of a patch, then interpolation methods
to find the point of intersection may require that we are
able to evaluate functions defining the vector field and the patch
boundary at the computed point outside the patch. If we check that
an integration step remains inside the chart containing the patch,
then we are able to apply standard algorithms. If a computed point
lies outside the chart, then remedial action like diminishing a step size
can be taken.
\vspace{1in}

\section*{An Example: Raibert's Hopper}

A partial implementation of the approach for
modeling hybrid dynamical systems presented
above has been accomplished in the {\bf dstool}
program \cite{back1}.  This Section describes
results for a particular control problem involving
the dynamical behavior and stability of
a ``hopping'' robot, constrained to move in a
single (vertical) dimension.
The study, both theoretical and experimental, of motion
and balance in legged robots
has a long history \cite{raibert1};  the model we
describe here was proposed recently
by Koditschek \cite{kod1} as a partial abstraction of a
physical machine built and tested
by Raibert \cite{raibert2}.

Raibert's robot consists of two main components,
a {\em body} which houses the control
mechanism and a compressible {\em leg} upon the base
of which all of the impact dynamics
of the robot with its environment occurs.  The leg is
constructed as a pneumatic cylinder
whose pressure is subject to closed feedback control
by opening and closing valves connected to two reservoirs
of compressed gas.
A schematic cartoon showing the important features
of the device relevant to Koditschek's
model is shown in Figure (\ref{ex1}).  To activate
the physical device, the robot is dropped
a short distance above a flat surface and, for some
values of the control parameters and after
an initial transient phase, the vertical motion of
the robot becomes periodic.  Indeed, this
behavior is remarkably stable under perturbation;
following a gentle impact, the motion
of the machine will autonomously return to its
periodic ``hopping'' state.

%
%

The dynamics of the hopping robot may be described
in terms of four distinct {\em phases}.  Let
$y$ denote the height of the robot body relative
to the surface above which the mechanism
is dropped.
The device is in its {\em flight} phase whenever the
leg is not in contact with the surface,
and the governing dynamics are presumed to be Newton's
equations for the one-dimensional
motion of a particle in a gravitational field with
constant acceleration, $g$.  During this
phase, the leg is kept at its fully extended length,
$l$, by a gas pressure inside the
pneumatic cylinder, $p_{init}$, maintained from one
of the gas reservoirs.  The robot enters
the {\em compression} phase as soon as the leg makes
contact with the surface, and the
weight of the body begins to compress the gas in the
pneumatic cylinder.  This compression
is modeled by a nonlinear spring, whose force is
inversely proportional to the robot height with a spring
constant, $\eta$, related to the initial gas pressure,
$p_{init}$.  Associated with the action
of the nonlinear spring is a simple mechanical damping
force with coefficient of friction denoted
by $\gamma$.  The {\em thrust} phase
begins at the point of maximal leg compression or,
equivalently, at the minimum value of the
$y$ coordinate achieved in the compression phase.  Let
$t_b$ denote the relative time at which
this minimum occurs.  At the onset of the thrust
phase, a valve opens to admit gas from the
second reservoir into the pneumatic cylinder at
a pressure $p_{th}$, which continues for a
\underline{fixed} period of time, $\delta$, and
exerts a constant force, $\tau$,
which accelerates the body upward.  The thrust
phase ends at the time
$t_b + \delta$ corresponding to a body height
  $y_{et}$.  At this point the valve closes,
and initiates the final {\em decompression} phase
where the robot body continues to move upward
under the action of the nonlinear pneumatic
spring.  The new nonlinear spring
constant, $\tau y_{et}$,  in effect for this phase
is a function of both the pressure of the gas injected
into the cylinder during the thrust phase
and the height at which the thrust phase was
completed.  Once the height of the body
reaches the length of the fully extended
leg,  the decompression phase ends and the
mechanism reenters the flight phase.  Figure (\ref{ex2})
illustrates the four phases
which comprise the behavior of the dynamical model.

%
%

The crucial step in preparing a hybrid dynamical
system for numerical simulation and analysis in {\bf dstool}
involves identifying the components of the model in
terms of the formalism described in the previous Sections.
For the hopping robot, it is natural to express the
governing dynamics as a two-dimensional
vector field on a phase space composed of four
patches, each patch corresponding to one of the phases
shown in Figure (\ref{ex2}).  Thus,
the index set, $\left \{1,2,3,4 \right \}$, corresponds
to the flight, compression, thrust and
decompression patches, respectively.  The vector field
is non-autonomous and dependent upon a six-dimensional
vector of parameters,
$p=\left (l,g,\eta,\gamma,\tau,\delta \right)$, whose
components were described earlier.
The definition of the patch regions, $U_\alpha$, are
derived by considering the simple required conditions
which must be imposed upon $(y,\dot{y},t)$.  The equations
of motion follow from application of Newtonian
force balance laws on each patch.  The model which results
may be expressed as,
\begin{equation}
  \ddot{y} = f(y,\dot{y},t; \alpha,p)
\end{equation}
where, $\alpha \in \left \{1,2,3,4 \right \}$, \hspace{.07in}
$(y,\dot{y},t) \in {\bf R}^{+}-\{0\} \! \times \!
 {\bf R}^{2}$ \hspace{0.03in}
and $p \in {\bf R}^{6}$.  Table (\ref{tab1}) summarizes
the important information required to characterize the
patches together with the equations of motion which
are in effect within each patch.  We note that $f_2$
and $f_4$ are singular whenever $y=0$ and all the
equations are physically unreasonable for $y<0$; thus,
we may take the open set
$V_\alpha = {\bf R}^{+}-\{0\} \! \times \! {\bf R}$ as
an appropriate chart for
each $\alpha \in \left \{1,2,3,4 \right \}$.

\begin{table}
\begin{center}
\begin{tabular}{||l|l|l||} \hline
\rule{0in}{.2in} Phase & \makebox[1in]{ $f$ }&
    \makebox[1.3in]{ Patch }  \\ \hline \hline
\rule{0in}{.2in}{\em Flight}     & $\; f(y,\dot{y},t;1,p) = -g$
  & $\; U_1 = \left \{ (y,\dot{y}) | y > l \right \} $   \\
\rule{0in}{.2in}{\em Compression
   }& $\; f(y,\dot{y},t;2,p) =\eta/y - \gamma \dot{y} -g$
  & $\; U_2 = \left \{ (y,\dot{y})  | y < l, \dot{y} < 0 \right \} $   \\
\rule{0in}{.2in}{\em Thrust}     & $\; f(y,\dot{y},t;3,p)
= \tau - \gamma \dot{y} -g$
  & $\; U_3 =  \{ (y,\dot{y})  | y < l, \dot{y} > 0,$ \\
  && \ \ \ \ \ \ \ \ \ \ \ \ \ \ \ \ \ \
  $ t_b < t < t_b + \delta \} $   \\
  \rule[-.1in]{0in}{.3in}{\em Decompression}& $\; f(y,\dot{y},t;4,p)
   = \tau y_{et}/y -\gamma \dot{y} -g$
  & $\; U_4 =  \{ (y,\dot{y}) | y < l, \dot{y} > 0,$ \\
  && \ \ \ \ \ \ \ \ \ \ \ \ \ \ \ \ \ \
  $ t > t_b + \delta  \} $    \\
\hline \end{tabular} \end{center}
\caption{\label{tab1} Definition of the vector field
components for Raibert's ``hopping'' robot.   }
\end{table}

Recall that the definition of a patch was the
positive open set for a collection of boundary
functions:  $h_{\alpha,i}(y,\dot{y},t;p) - C_{\alpha,i} > 0$.  These
boundary functions are obtained from
the set notation of Table (\ref{tab1}) by
translating the definition of intervals
into statements involving functions of the
independent variables $y$, $\dot{y}$ and $t$ and target
values, $C_{\alpha,i}$.  It is important to note that
this translation is not unique.  For example,
the relations
$y>l$ and $y-l>0$ are equivalent expressions regarding
the same mathematical object.  {\bf Dstool}
allows the analyst to take advantage of repetition in lists
of boundary functions required to define a
hybrid dynamical system.  If a function, $h$, can be used in
several instances to express boundary functions,
with or without different target constants
($C_{\alpha,i}$), a {\em single block of code} can be used
to define the boundary condition in each case.  Thus, the
user may minimize the work required to install
the model by reuse of the definitions of boundary
functions.  As a concrete illustration,
Table (\ref{tab2}) provides a list of boundary functions
and target constants for the legged robot, corresponding
to the open sets defined in Table (\ref{tab1}).  While there
are nine boundary conditions required, only
six {\em C}-language software functions must be written
to accomodate this list.  Admittedly, for the
purposes of this example the savings is small but in
larger dynamical systems, especially those composed of
many similar subcomponents, the advantages gained in code
size and error reduction may be significant.

\begin{table}
\centering
\begin{tabular}{||l|cc|l|c||} \hline
      & Index & Index & & \\
Phase & Key $(\alpha)$ & $(i)$ & \makebox[1.in]{
$h_{\alpha,i}$} &
\makebox[1.in]{ $C_{\alpha,i}$}  \\ \hline \hline
\rule{0in}{.2in}{\em Flight}     & 1 &
1 & $h_{1,1}(y,\dot{y},t;p) = y$ & $l$  \\
\rule{0in}{.2in}{\em Compression}
 & 2 & 1 & $h_{2,1}(y,\dot{y},t;p) = -y$ & $-l$  \\
 &   & 2 & $h_{2,2}(y,\dot{y},t;p) = -\dot{y}$ & $0$  \\
\rule{0in}{.2in}{\em Thrust}
 & 3 & 1 & $h_{3,1}(y,\dot{y},t;p) = -y$ & $-l$  \\
 &   & 2 & $h_{3,2}(y,\dot{y},t;p) = \dot{y}$ & $0$  \\
 &   & 3 & $h_{3,3}(y,\dot{y},t;p) =
  -t$ & $-(t_b + \delta)$  \\
\rule{0in}{.2in}{\em Decompression}
 & 4 & 1 & $h_{4,1}(y,\dot{y},t;p) = -y$ & $-l$  \\
 &   & 2 & $h_{4,2}(y,\dot{y},t;p) = \dot{y}$ & $0$  \\
 &   & 3 & $h_{4,3}(y,\dot{y},t;p) = t$ & $(t_b + \delta)$  \\
\hline \end{tabular}
\caption{\label{tab2} Boundary functions for the patches
 used in the model for Raibert's ``hopping'' robot.   }
\end{table}

Together Tables (\ref{tab1}) and (\ref{tab2}) contain the
information required to install the hybrid model
for Raibert's legged robot into the {\bf dstool}
program.  For details
regarding the installation of dynamical systems
into {\bf dstool} see \cite{back1,gucken1}.
For this example, eight {\em C}-language procedures that
provide information required by the program must be
written by the {\bf dstool} user.
The routines may be described as follows:

\begin{itemize}
  \item \makebox[1in][l]{ \underline{\bf model\_init()}: }
       \parbox[t]{3.4in}{ This procedure is called when the
       dynamical system is first initialized to provide
     basic information such as the
     dimension and names of the variables, Jacobian
     information, and plotting ranges. }

  \item \makebox[1in][l]{ \underline{\bf info()}: }
       \parbox[t]{3.4in}{ A routine is required to
       inform the data structure object manager of the name
of each procedure used to define the components of the atlas and
transition functions. }

  \item \makebox[1in][l]{ \underline{\bf equations()}: }
  \parbox[t]{3.4in}{ The vector field equations,
  $f(y,\dot{y},t;\alpha,p)$ are specified
in this procedure for each chart in the atlas. }

  \item \makebox[1in][l]{ \underline{\bf patch\_def()}: }
 \parbox[t]{3.4in}{ Information which defines each patch, $U_\alpha$,
is provided by this routine including  target constants and
tolerances to be used.}

\item \makebox[1in][l]{ \underline{\bf bndry\_fncts()}: }
\parbox[t]{3.4in}{ Each boundary function which participates
in the definition of the patches contained in the
atlas are specified by
this procedure. }

\item \makebox[1in][l]{ \underline{\bf trans\_fncts()}: }
\parbox[t]{3.4in}{ The set of transition functions
to be applied
at the various boundaries of the patches are collected
together in this routine.}

\item \makebox[1in][l]{ \underline{\bf traj\_init()}: }
\parbox[t]{3.4in}{ At the beginning of a trajectory
segment, the user may require certain conditions
or parameters be checked and used to set characteristics
of the model before
propagation begins.  This routine provides this facility.  }

\item \makebox[1in][l]{ \underline{\bf size()}: }
\parbox[t]{3.4in}{ One important responsibility
expected of the data structure object manager
is the allocation and deallocation of system
memory.  To fulfill this task,
the manager must be provided with information
concerning the size of each
element in the atlas.  This routine may
be called on demand to provide these
dimensions. }
\end{itemize}

Once these routines are installed in {\bf dstool},
the program may be used to propagate trajectories
using standard versions of integration
algorithms.  Collections of trajectories generated
using different
initial conditions may be displayed together to form
phase portraits, which may be used to eluciudate
global structures and bifurcations which vary with
system parameters.  Figure (\ref{ex1}) displays a
periodic orbit for the hopping robot.  The upper plot
shows the phase portrait corresponding to thrust force, $\tau$,
of $41.86$ units.  The trajectory sense indicated is clockwise,
with the robot height, $y$, displayed on the abscissa
and the velocity along the ordinate.  The four mechanical
phases which contribute to the cycle are clearly evident:
The flight phase occurs above the point $y=l$, followed by
the compression of the pneumatic spring, a short
period of thrust and finally the full expansion of
the robot leg.
The lower plot in this Figure displays the height of the
robot body as a function of time.
Each revolution of the limit cycle corresponds to a single
robot ``hop'',  and the maximum height attained is
constant. This is an example of a {\em  attracting}
limit cycle; if initial conditions are chosen which do
not lie on the periodic orbit and integrated forward
in time the resulting orbit will approach the cycle
asymptotically in the limit.  Clearly, if one presumes
this model to
represent the dynamics of the physical device, the
remarkable
stability of Raibert's hopper may be interpreted
as a direct consequence of the large domain of
attraction for this limit cycle.

%
%

As the thrust parameter, $\tau$, is increased
a {\em period-doubling} bifurcation of this periodic
orbit occurs.  Figure (\ref{ex2}) shows an example
of a period-2 orbit which results at a value of
$\tau=68.5$. Again, the upper plot shows a phase portrait with
$y$ and $\dot{y}$ displayed along the axes.  Notice that a
single limit cycle consists of two components, each
component involving one compete set of the four
mechanical phases.  Moreover, one subcomponent of the
limit cycle
may be characterized by the observation that the
maximum height attained during its flight phase is only about half
that of the second subcomponent.  A physical device
displaying this behavior would take a short hop, followed
by a higher hop and repeat this sequence
indefinitely.  Once again, this periodic orbit can be shown to be
stable with a large domain of attraction, so that
if such a physical device were perturbed - bumped - after an
initial transient period it would return to
this two-stage hopping behavior.  The presence of this
period-doubling
bifurcation has been previously reported by Vakakis
and Burdick \cite{vav1} in numerical studies and Koditschek
\cite{kod1} in simplified analytical approximations.  These
period-two trajectories  may also possibly be confirmed by
experimental observations by Raibert, who coined the
term ``limping gait'' to characterize the corresponding  motions.

\section*{Conclusions}
We have presented a methodology for studying dynamical
systems whose phase spaces involve both continuous and
discrete components, representing a large and important
class of models.  A software program for studying
dynamical systems has been enhanced to accomodate
hybrid models of the genre described, and this tool has
been shown to reproduce established behavior for a
selected problem in robotics.  Further development
of software tools for investigating hybrid systems
is underway and this initial implementation supports
these efforts in two ways.  First, this software can
serve as a
platform for evaluating the methodology presented
here on problems of current research interest.
The representation of hybrid systems may then
be refined and extended.  Second, the study of
such systems will
undoubtedly require new numerical algorithms which
can be developed and carefully tested using programs
based on the initial efforts described in this
paper.  We briefly describe two directions of current
activity.

In sophisticated hybrid dynamical systems, the geometry
of the various components which comprise the phase space and
boundary sets becomes important.  We are interested in
extending the model described in the first Section
to accomodate the complex relationships which result
near the boundaries of the patches.  We foresee the need to
further decompose the patch boundaries into sets where
the action of the transition functions may vary,
according to requirements dictated by their relative
geometry or significance as an entity in the model.   Within
this decomposition, conditions for discrete state changes
might result from combinations of Boolean functions
based on both discrete state information and phase
space coordinates.  Thus, the appropriate action of the transition
functions near the intersection of patch boundaries or for a
trajectory nearly parallel to a boundary, for example, can be
discriminated.

The body of literature describing numerical algorithms
for vector fields represent by Equation (\ref{eq1}) is
large and growing rapidly.  In some cases, adapting these
procedures for use with hybrid dynamical systems
should require little more than safeguarding their
behavior near patch boundaries; locating fixed points
using Newton-type methods is such an example.  In other
cases, extension is more difficult.  Consider, for example
the problem of approximating a periodic orbit in the phase
space of a hybrid vector field that spans more than
a single patch.  Simulation may be used,  as it was
in the Example problem above, to search for an initial
condition in the phase space that returns to the starting
condition after some fixed period.
Another method widely used for Equation (\ref{eq1}) is to
solve an appropriate two-point boundary value problem using
parallel shooting or collocation methods.  However, in the
case of a hybrid system on multiple patches, this
naturally yields a multipoint boundary-value problem without
the smoothness properties often assumed at the interior
points.  Thus, the extension and development of numerical
procedures for hybrid systems poses new and interesting
problems for analysis.

\section*{Acknowledgements}

The authors acknowledge Dr. Anil Nerode of
Cornell University and the Mathematical
Sciences Institute whose suggestions provided
the original impetus for this work.
We also thank Dr. Daniel Koditschek at Yale
University for a number of valuable
discussions concerning the dynamics of Raibert's hopper.
This research was partially supported by the National
Science Foundation and the Defense Advanced
Research Projects Agency.

%
%

\clearpage

\noindent \underline{\bf \Large Figure Captions:}
\vspace{.25in}


\setcounter{figure}{0}


\refstepcounter{figure} \label{exa}
\noindent Figure~\ref{exa}:  Schematic diagram
illustrating the basic components of a
hybrid dynamical system.
\vspace{.15in}

\refstepcounter{figure} \label{ex1}
\noindent Figure~\ref{ex1}:   Schematic of a simplified
model for Raibert's hopping robot.
\vspace{.15in}

\refstepcounter{figure} \label{ex2}
\noindent Figure~\ref{ex2}:   Sequence of phases which
comprise the robot's hopping dynamics.
\vspace{.15in}

\refstepcounter{figure} \label{ex3}
\noindent Figure~\ref{ex3}:  Periodic ``hopping'' of Raibert's robot.
Figure (4a) shows
the two-dimensional phase portrait,
the coordinate $y$ along the ordinate and
$\dot{y}$ on the abcissa.  The height of
the body, $y$, plotted as a function of time
is shown in Figure (4b).
\vspace{.15in}

\refstepcounter{figure} \label{ex4}
\noindent Figure~\ref{ex4}:  Limping gait motion of
Raibert's robot which results
from a period doubling bifurcation.
Figure (5a) shows
the two-dimensional phase portrait,
the coordinate $y$ along the ordinate and
$\dot{y}$ on the abcissa.  The labels $F,C,T$
and $D$ denote the flight, compression,
thrust and decompression phases of the mechanical device.  The
height of the body, $y$, plotted as a function of time
is shown in Figure (5b).

\end{document}